\documentclass[prb,preprint,amssymb]{revtex4-1} 

\usepackage{amsmath}  
\usepackage{amsfonts} 
\usepackage{graphicx} 

\begin{document}


\title{Square-Well Approximation for the Anharmonic and the Double-Well Oscillators}

\author{B.P.Mahapatra}
\email{bimal58.mahapatra@gmail.com} 
\altaffiliation[permanent address: ]{1065/3486,Jagmohan Nagar, Bhubaneswar 751030,India} 
\affiliation{School of Physical Sciences, National Institute of Science Education and Research (NISER), Bhubaneswar 751005, India}

\author{N. B. Pradhan}
\email{noubeharypradhan1963@gmail.com}
\affiliation{Department of Physics, G.M.(Autonomous)College, Sambalpur 768004, India}


\date{\today}

\begin{abstract}
A novel general approximation scheme (NGAS) proposed earlier \cite{ar1,ar2} is applied to the problem of  the quartic anharmonic (QAHO)  and the double-well oscillator (QDWO) in quantum theory by choosing the infinite square-well potential in one dimension as the input approximation. The leading order (LO) results obtained for the energy eigen-values are uniformly accurate to within a few percent of the exact results for \emph {arbitrary} values of the quartic-coupling: $\lambda > 0$ and for \emph{all} the energy levels: $n_s\geq0$. These results are shown to be non-perturbative in the LO and reproduce the known analytic and scaling properties of energy as a function of the coupling $\lambda$  and the level-index: $n_s$. The LO-results are further improved in accuracy by including the perturbative-correction at the next non-trivial order of an improved perturbation theory (IPT) based upon NGAS. The method can be trivially extended to other cases of higher anharmonicity. 

\end{abstract}
\maketitle 

\section{Introduction} 

 The anharmonic oscillator (AHO) and the double-well oscillator (DWO) are among the simplest of the interacting quantum systems which find extensive application in various areas\cite{ar1} of physics and chemistry. These include areas as diverse as atomic and molecular physics, quantum chemistry, condensed matter physics, particle physics, statistical physics, quantum field theory and cosmology. However, exact analytical solutions are not available for the above systems. One must therefore resort to approximations to solve these cases. For this reason, there is a lot of theoretical/mathematical interest in testing and improving upon the existing approximation schemes applied to the AHO/DWO systems. These investigations include perturbation theory, variational methods, path-integral approximation, variation-perturbation-methods, summability-methods , Pade-approximation etc. Consequently, there is a vast amount of literature accumulated\cite{ar1} on the subject demonstrating the perennial interest in these systems.  

       The AHO in one dimension, with quartic-, sextic- and octic-anharmonic- interaction, as well as, the DWO systems with quartic and sextic-anharmonicity were earlier investigated\cite{ar2,ar3} by the present authors in a novel general approximation scheme (NGAS). This scheme, by construction, is potentially capable of providing accurate approximation for any {\it general} interaction of {\it arbitrary strength} in quantum theory. The basic input to the scheme is the choice of a suitable approximating-Hamiltonian ${H_{0}}$ , which is required to be exactly solvable, but which  involves certain adjustable (variational) parameters, $\alpha_{i}$. The scheme is implemented by imposing the constraint that the quantum average (QA) of the original Hamiltonian $H$ be equal to that of the approximating Hamiltonian ${H_{0}}$ with respect to any (arbitrary) eigen-state of the latter. In the next step of implementation of the scheme, the (variational) minimization of this QA with respect to the free-parameters:$~ \alpha_{i}$ of the model,completely determine the latter.The evaluation of the energy eigen-value of the reference(input) Hamiltonian then determines the leading order(LO) result in NGAS.
       
           It has been demonstrated \cite{ar2,ar3} that this simple procedure automatically builds into the approximating Hamiltonian ${H_{0}}$, the nonlinearity and other symmetries of the original Hamiltonian, thereby guaranteeing accurate results for \emph{arbitrary} coupling strength, $\lambda$ and for $\emph{arbitrary}$ excited states of the system ${\it even}$ in the LO.  Furthermore, the LO-results thus obtained, are seen to be essentially non-perturbative, yet improvable order-by-order in an improved perturbation theory (IPT) defined in NGAS. This IPT is distinct from the conventional perturbation method in several respects. In particular, the IPT is \emph{not restricted to `small-coupling-expansion'}. 
           
     In the earlier investigations \cite{ar2,ar3}, the simple harmonic-approximation (SHA) was made as the natural choice in selecting the input Hamiltonian, $H_{0}$ as that for the simple harmonic oscillator (SHO) but suitably  generalized to include variable parameters corresponding to the frequency, an over-all energy- shift and an appropriate choice of the ground-state configuration (to take into account the symmetry breaking mechanism in case of the DWO). With these inputs, excellent results for the energy-spectrum in all the above cases were \emph{uniformly} achieved for arbitrary coupling strength of anharmonicity $\lambda$, as well as, for arbitrary excitation levels,`n' \emph{even} in the LO. In addition, this study \cite{ar2,ar3} also revealed important insight into the structure of the interacting vacuum, the instability of the perturbative-ground state. Furthermore, consistency with the results of super-symmetry were demonstrated wherever applicable. Similarly, in the domain of quantum field theory, NGAS has been applied  \cite{ar4} to the case of $\lambda\phi^{4}$ theory , with the choice of approximating Hamiltonian as that for the free-hermitian-scalar field (but with adjustable `mass' and `shifted' field). In this case, the standard results of the Gaussian-approximation\cite{ar5} were reproduced \cite{ar4} in the LO including the non-perturbative renormalization of the `mass' and `coupling strength'.
     
      In the current investigation, we exploit the freedom of choice of the input approximating Hamiltonian in NGAS to choose the same for the infinite-square-well (ISW)-potential.One important objective in selecting the ISW-potential as the input is the possible pedagogical interest as well as the simplicity of this choice. It is well-known that the ISW-potential constitutes one of the simplest systems admitting exact analytical solution. As such, it is included in any standard course of introductory quantum mechanics and introduced to the learner fairly early in the subject. The establishment of an approximation connecting the AHO/DWO to the ISW-case, may very well be considered as an illustrative example of application of a standard text-book topic to advanced research.
      Admittedly, however, the ISW- approximation is perhaps the crudest among possible choices. This is due to the fact that the system subjected to the ISW potential propagates \emph{freely} between the infinite walls, in stark contrast to the actual situation for the AHO/DWO. Nevertheless, we choose it on purpose here, in order to test the robustness and tolerance of NGAS to the crudest possible   input approximation. Moreover, as a by-product of this study, it is possible to obtain an approximation for the celebrated case of the SHO by simply setting the anharmonic- coupling strength to zero in the AHO-Hamiltonian, thus gaining further insight into the accuracy of the approximation. 
            
      The paper is organized as follows. In the next section, we recall the main steps for the general formulation of NGAS. In Section-III, we demonstrate the method by applying it to the case of the quartic-AHO choosing the ISW-Hamiltonian as the input. The LO-results for the energy spectrum are obtained and compared with the results from other calculations. The non-perturbative aspects and the analytic structure of energy as a function of the quartic-coupling are also discussed. In Section-IV , analogous results for the quartic- DWO and the SHO are presented. We outline the method of the improved perturbation theory (IPT) based in NGAS in Section-V , compute the next-order correction to the energy and show how the LO-results could be further improved by  this correction. Finally, we conclude in Section-VI with a summary and discussion of the results.
    \section{BASIC FEATURES OF THE GENERAL FORMULATION OF NGAS}
 Consider a quantum system described by the Hamiltonian,$H\equiv H(x,\hat{p};\lambda)$ which is to be solved to obtain the energy-spectrum and the eigen states. Here we consider a one-dimensional system for simplicity and denote by $\lambda$, the {\it generic} strength of interaction in $H$. The first step in implementation of NGAS consists of the choice of a {\it suitable} approximating Hamiltonian,$ H_0(x,\hat p,\{\alpha_i\})$ which is {\it exactly solvable} and which, is chosen to depend upon a set of the ${\it free}$ (adjustable) parameters, $\{\alpha_i\}$. We denote the eigen-value equation for $H_0$ as:
\begin{eqnarray}
H_0|\phi_n(x)\rangle~=~\mathcal{E}_n|\phi_n(x)\rangle;
\label{f1}
\end{eqnarray}
 and assume the normalization of the eigen-states expressed in standard notation as: $\langle\phi_n(x)|\phi_n(x)\rangle~=~1$.

The next step in the implementation of NGAS is the imposition of the ${\it constraint}$:
\begin {eqnarray}
\langle\phi_n(x)|H|\phi_n(x)\rangle~=~\langle\phi_{n}(x)|H_{0}|\phi_{n}(x)\rangle \nonumber \\
~~~~~~~~~~~~~~~~~~~~~~~~~~~~~~~~~~~~\equiv~\mathcal E_{n}(\{\alpha_{i}\},\lambda)
\label{f2}
\end{eqnarray}
In what follows, eq.(1) will be referred as the {\it ``Constraint of Equal Quantum Average (CEQA)''}.
  In the third step, the variational minimization of  $\mathcal E_n({\alpha_{i}},\lambda)$,
  with respect to the free-parameters, as given below
  \begin{eqnarray}
  \frac{\partial \mathcal E_{n}}{\partial\alpha_{i}}~=~0,
  \label{f3}
  \end{eqnarray}
  determines the adjustable parameters to their {\it optimal} values:
\begin{eqnarray}
{\alpha_{i}}~=~{\beta_i(\lambda,n)}
\label{f4}
\end{eqnarray}
   Eq.(3) is referred as the `Condition of Variational Minimization(VMC)' in the following. The leading order(LO) result for the energy eigenvalues are then obtained by substitution of Eq.(\ref{f4}) in Eq.(\ref{f2}) and given by:
\begin{eqnarray}
{\mathcal E_n({\beta_{i}},\lambda)}~\equiv~{ E_n^{LO}}
\label{f5}
\end{eqnarray}
  In the final step of implementation of NGAS, the LO-results can be systematically improved further by an improved perturbation theory (IPT) as follows\cite{ar2,ar3}:\\
   (i) The unperturbed Hamiltonian is chosen to be $H_0$ as given by Eq.(\ref{f2}) but incorporating Eq.(\ref{f4});
 (ii) Then the {\it natural} choice for the `perturbation-Hamiltonian' becomes:  
\begin{eqnarray}
{H^{'}}~=~{\it H-H_{0}}
\label{f6}
\end{eqnarray}  
It may be noted that a direct consequence of CEQA, Eq.(2) is the result:
\begin{equation}
\langle\phi_n(x)|H^{'}|\phi_n(x)\rangle~=~0.
\label{f7}
\end{equation}
Other features of IPT are discussed in Section-V. In the next Section we discuss the application of NGAS to the AHO/DWO system in the ISW-approximation.

\section{The ISW-approximation for the Quartic-AHO and the SHO}
\textbf{A.The Quartic AHO}\\
 The quartic-AHO is defined by the Hamiltonian in the following \emph{dimension-less} form:
\begin{eqnarray}
{H}~=~{\frac{1}{2}~p^2}~+~{\frac{1}{2}~gx^2}~+~{\lambda{x^4}}
\label{f8}
\end{eqnarray}
where, $g>0$ and $\lambda > 0$. The input-Hamiltonian for the ISW-potential is defined by: 
\begin{eqnarray}
{H^{SW}_{0}~}=~{\frac{1}{2}~p^2}~+~{V_{SW}}
\label{f9}
\end{eqnarray}
where the potential, $V_{SW}$ is given by:   
\begin{eqnarray}
{V_{SW}(x)}~=~\left\{\begin{array}{rc}
\infty & \mbox{$|x|\ge a$}\\
h & \mbox{$~~|x|<~a~$},\\
\end{array}
\right.
\label{10}
\end{eqnarray}
Note that the two free-parameters which characterize the ISW-potential are the $\it `width' (~=~ 2a)$ and the $\it `depth'~(= h)$. Note also that the potential is chosen to be symmetric under space-inversion:
$ V_{SW}(-x)~=~ V_{SW}(x)$ in order to preserve the same symmetry of the original Hamiltonian given by Eq.(8).
The eigenvalue equation for $\it H_0^{SW}$ as given below
\begin{eqnarray}
H^{SW}_0~\phi^{SW}_n(x)~=~\mathcal{E}_n~\phi^{SW}_n(x),
\label{f11}
\end{eqnarray}
is easily solved subject to the boundary-condition:
\begin{eqnarray}
\phi^{SW}_n(x)~~~=~0  ~~~~~\textrm{for}~~ |x|\geq{a}, 
\label{f12}
\end{eqnarray}
which ensures the physical requirement of absolute confinement of the system between the (infinite)potential barriers. For $~~|x|< a,$ the normalized eigen-functions vanishing on the boundary of the potential-well are given by:
\begin{eqnarray}
\phi^{SW}_n(x)\equiv\phi^{(-)}_{n}(x)~=~\frac{1}{\sqrt{a}}~sin\left(\frac{n\pi x}{2a}\right);~~\textrm{for}~~ n=2,4,6,8...; 
\label{f13}
\end{eqnarray}
and,
\begin{eqnarray}
\phi^{SW}_n(x)\equiv\phi^{(+)}_{n}(x)~=~\frac{1}{\sqrt{a}}~cos\left(\frac{n\pi x}{2a}\right);~~\textrm{for}~~ n=1,3,5,7.... 
\label{f14}
\end{eqnarray}   
In the above equations, the $(\pm)$ super-scripts correspond to the even(odd)-parity solutions. The energy eigen-values are trivially obtained by solving the Schr\"{o}dinger equation and given by
\begin{eqnarray}
{\mathcal{E}_{n}}~=~{h}~+~\frac{n^2\pi^2}{8a^2}
\label{f15}
\end{eqnarray}
  The next task is the determination of the two adjustable parameters,$\it{ `a'}$ and $\it{ `h'}$. The width-parameter $`a'$ can be determined by the variational-minimization of ${<H>}$. Here, we use the notation:${<\hat {A}>}$  to define the quantum-average /expectation-value of the operator, ${\hat {A}}$ as given below:
\begin{eqnarray}
\langle {\hat A}\rangle ~\equiv~\int_{-a}^{+a}dx~{\phi_n^{\ast}(x)~\hat A~\phi_n(x)}
\label{f16}
\end{eqnarray}
  The evaluation of $<H>$ using the above definition,eq.\eqref{f16}is straight-forward. Noting that:
  \begin{eqnarray}
  \langle H\rangle~=~ \langle\frac{1}{2}~p^2\rangle~+~\langle\frac{1}{2}~gx^2\rangle~+~\langle\lambda x^4\rangle,
  \label{f17}
\end{eqnarray}
each term in eq.(17) can be computed easily by exploiting parity-invariance, which forbids parity-changing transitions,i.e. $<\phi_n^{+}|H|\phi_n^{-}>~=~0~=~<\phi_n^{-}|H|\phi_n^{+}>$. The result is given below:
 \begin{eqnarray}
 \langle H\rangle~=~ \left( \frac{n^2\pi^2}{8}\right)\left(\frac{1}{a^2}\right)~+~\left(\frac{g}{6}\right)c_n a^2~+~\lambda a^4\left(\frac{1}{5}-\frac{4c_n}{n^2 \pi^2}\right)
  \label{f18}
 \end{eqnarray}
where,
 \begin{eqnarray}
c_n~\equiv~1-\left(\frac{6}{n^2 \pi^2}\right),~~{n= 1,2,3,4,...}.
 \label{f19}
 \end{eqnarray}
 The minimization of the expression for $<H>$ as given above, with respect to $u\equiv~\left(1/a^2\right)$ leads to the following equation:
 \begin{eqnarray}
u^3-P\left(g,n\right)u-Q\left(\lambda,n\right)~=~0,
 \label{f20}
 \end{eqnarray}
 where,
\begin{eqnarray}
P\left(g,n\right)~\equiv~\left(\frac{4}{3}\right)\left(\frac{gc_n}{n^2 \pi^2}\right),
\label{f21}
\end{eqnarray}
\begin{eqnarray}
Q\left(\lambda,n\right)~\equiv~\left(\frac{16\lambda}{n^2 \pi^2}\right){\left(\frac{1}{5}-\frac{4 c_n}{n^2 \pi^2}\right)}.
\label{f22}
\end{eqnarray}
The real, positive root of eq.(20) is required on physical grounds. This is given by,
\begin{eqnarray}
u=\left\lbrace{\left(\frac{8\lambda}{n^2 \pi^2}\right)\left(\frac{1}{5}-\frac{4 c_n}{n^2 \pi^2}\right)}\right\rbrace^\frac{1}{3}\left[\left\lbrace 1+\sqrt{1-\rho}\right\rbrace^\frac{1}{3}+\left\lbrace 1-\sqrt{1-\rho}\right\rbrace^\frac{1}{3}\right],
\label{f23}
\end{eqnarray}
where
\begin{eqnarray}
\rho~\equiv~\left(\frac{4}{27}\right)\left(\frac{P^3}{Q^2}\right).
\label{f24}
\end{eqnarray}
Substitution of Eq.(20) in Eq.(18) and following Eq.(2)and Eq.(5), one obtains the following simple expression for the energy eigen-values in the LO: 
\begin{eqnarray}
E_n^{LO}~=~\left(\frac{3n^2\pi^2}{16}\right)u~+~\left(\frac{gc_n}{12}\right)\left(\frac{1}{u}\right),
\label{f25}
\end{eqnarray}
where, $u$ is given by eq.(23). The remaining parameter, $`h'$ can then  be determined by substitution of eq.(25) in eq.(15) as follows:
\begin{eqnarray}
h~=~\left(\frac{n^2\pi^2}{16}\right)u~+~\left(\frac{gc_n}{12}\right)\left(\frac{1}{u}\right)
\label{f26}
\end{eqnarray}
At this point, several observations are in order:\\
(i) note that the free parameters of the input-Hamiltonian,i.e. $`a'$ and $`h'$ acquire functional dependence on  $`\lambda'$ and $`n'$  through Eq.(23) and Eq.(26). This in turn , implies that the input-Hamiltonian,$H_0^{SW}$ (see,eq.(9) also becomes a function of $\lambda$ and $ n $. An obvious consequence is that the eigen-functions of $H_0^{SW}$ corresponding to different eigen-values become non-orthogonal,i.e.
\begin{equation}
(\phi_m(x),\phi_n(x))~\equiv~\int_{-a}^{+a}dx~{\phi_m^{\ast}(x)\phi_n(x)}\neq 0~for ~m\neq~n,
\end{equation}     
(ii) It is seen from Eq.(23) and Eq.(25) that the energy in the LO is a non-analytic function of $\lambda$ at the origin. Hence, the results expressed in these equations are \emph{not} accessible to ordinary perturbation theory as a power-series expansion in $\lambda$. In this sense, the LO-results are \emph{non-perturbative}.\\
(iii) The cube-root singularity at $\lambda=0$ and the branch-point-structure in the $ complex~\lambda-plane$ as given in Eq.(25) are in conformity with  rigorous derivation \cite{ar6} of the analytic structure of the energy of the quartic-AHO in the coupling strength-plane using sophisticated tools of complex-analysis. \textit{It is  remarkable that, the correct analytic-structure in $\lambda$  arises here  as a simple consequence of the CEQA and the CVM of NGAS}( see, Eq.(2) and Eq.(3)).

It needs to be emphasized here that the observed features noted above under (i)-(iii) also hold in the analogous case of the simple-harmonic approximation \cite{ar2,ar3} applied to the same examples of the AHO/DWO. This fact demonstrates that the noted features are the  \emph{inherent} properties of the scheme, NGAS, arising \emph {independent} of the choice of the input Hamiltonian.

We show in Table-I below, the LO-results for the energy-levels of the quartic-AHO at sample-values of the oscillator-level-index $ n_s $ and the quartic-coupling $\lambda$ for $g~=~1$. ( Note that the ISW-level-index and the oscillator-level-index differ by one unit, i.e. $ n_s~=~ n-1$, for  n = 1,2,3,4,.. ). Also shown in this Table are the energy levels, $E_n^{(2)}$ which include 2nd-order corrections in IPT (see, Section-V) as well as,earlier results from ref.(7) for comparison. It is seen from this tabulation that the LO-results are uniformly accurate to within $\sim (2-12)~\%$ of the standard results over the full range of $n_s$ and  $\lambda$. It is further seen that the accuracy of the LO-approximation increases with the increase of the level-index, $n_s$.

The energy-eigenfunctions in the LO are those as given by Eq.(13) and Eq.(14) but with the width-parameter,$`a'$ appearing in these equations now becoming a function of $\lambda$ and $n$ in accordance with Eq.(23).

We next discuss the LO-approximation for the SHO in the next sub-section.\\
\textbf{B.~~The SHO}\\
The Hamiltonian of the SHO is simply obtained from that of the AHO, Eq.(8) by setting the quartic-coupling $~\lambda$ to zero and given by
\begin{eqnarray}
{H^{SHO}}~=~{\frac{1}{2}~p^2}~+~{\frac{1}{2}~gx^2.} 
\label{f28}
\end{eqnarray}
The exact analytic results for this system are,of course,well known and given in any standard text on quantum mechanics. There is therefore no need for any approximation for this celebrated example. However, these `exact'  results provide a further test of the accuracy and efficacy of the ISW-approximation in NGAS. It is purely in this context that the SHO is discussed here as a particular case ($\lambda= 0 $) of the AHO. The substitution: $\lambda=0$ in Eqs.(17-20) lead to the following corresponding results for the  SHO:
\begin{eqnarray}
{u=\sqrt{P},}
\label{f29}
\end{eqnarray}
\begin{eqnarray}
E_n^{LO}|_{SHO}~=~\left(\frac{n^2\pi^2}{8}\right)u~+~\left(\frac{gc_n}{6}\right)\left(\frac{1}{u}\right).
\label{f30}
\end{eqnarray}
The `depth' parameter for the case of the SHO is given by,
\begin{equation}
h_{SHO}~=~\left(\frac{gc_n}{6}\right)\left(\frac{1}{u}\right).
\end{equation}
The LO-approximation for the energy-levels following from Eqs.(29-30) are tabulated in Table-II for typical values of the level-index, $~n_s$ and for $~g~=1$ along with the results after inclusion of the 2nd-order correction in IPT ( discussed in Section-V ). The accuracy of the approximation for both the cases with respect to the exact analytic result are also shown in the same Table. In this context, it may be interesting to obtain an $asymptotic$ estimate for the accuracy of the LO-result given by Eq.(30) as compared to the $`exact'$ result. This is given by
\begin{equation}
\lim_{n_s\to\infty}\left(\dfrac{E_n^{LO}}{E_n^{exact}}\right)=\left(\dfrac{\pi}{2\sqrt{3}}\right)\simeq~0.9069,
\end{equation} 
which corresponds to an error of approximately $9.31\%$ . At finite values of $n_s$ the errors are of the same order of magnitude.However, the inclusion of the 2nd-order correction in IPT significantly improves the accuracy of approximation.

The next Section is devoted to the discussion of the ISW-approximation for the quartic-DWO.
\section{The ISW-Approximation for the quartic-DWO}
The Hamiltonian for the quartic-DWO is obtained from that of the AHO, Eq.(8) by a change of sign of the quadratic coupling, $~g~\rightarrow~-g$  and given by
\begin{equation}   
{H^{DWO}}~=~{\frac{1}{2}~p^2}~-~{\frac{1}{2}~gx^2}~+~{\lambda{x^4};~~~~~ \lambda~,g > 0}
\end{equation}
Apart from the various applications of the DWO as discussed earlier, there is considerable theoretical/mathematical interest in the system. In particular, the instability at $\lambda=0$( due to the non-existence of a physical ground state) prevents the application of the na\"{\i}ve perturbation theory. In some versions of modified-perturbation theory it has been established\cite{ar8} that power series-expansion in  $\lambda$ is not even Borel-summable. However, in the context of NGAS, the ISW-approximation is routinely applicable to the DWO-case as well, by merely a change of sign of the quartic-coupling:   $g\rightarrow~-g$ in the corresponding formulae for the AHO. In particular, the equation analogous to Eq.(20) for the DWO now becomes:
\begin{equation}
u^3~+~P\left(g,n\right)u-Q\left(\lambda,n\right)~=~0,
\end{equation}
leading to the $physical$ solution (analogous to Eq.(23)) given by
\begin{equation}
u=\left\lbrace{\left(\frac{8\lambda}{n^2 \pi^2}\right)\left(\frac{1}{5}-\frac{4 c_n}{n^2 \pi^2}\right)}\right\rbrace^\frac{1}{3}\left[\left\lbrace \sqrt{\rho+1}+1\right\rbrace^\frac{1}{3}-\left\lbrace \sqrt{\rho+1}-1\right\rbrace^\frac{1}{3}\right],
\end{equation}
The results for the energy in LO are therefore given by:
\begin{equation}
E_n^{LO}|_{DWO}~=~\left(\frac{3n^2\pi^2}{16}\right)u~-~\left(\frac{gc_n}{12}\right)\left(\frac{1}{u}\right),
\end{equation}
However, several authors \cite{ar7} have found it convenient to measure the energy of the DWO from the $\emph{bottom}$ of the (symmetric)double-well, which means that a term equal to $1/{16\lambda}$ be added to formula, Eq.(36). We denote this quantity as:~$E_n^{\textrm{ref}}\equiv~E_n^{LO}+\left(1/{16\lambda}\right)$.

In Table-II, we present the results  for  $E_n^{\textrm{ref}}$ along with results corrected to include 2nd-order perturbation effects in IPT ( see, Section-V) for a range of values of $\lambda$ and $ n_s $. The relative accuracy of these results with respect to those from earlier computation are also given in this Table. It can be seen from this tabulation that the ISW-approximation uniformly reproduces the standard results to within $\sim{(2-10)~\%}$ for the case of LO and to within $\sim{(0.5-7.5)~\%}$ with the inclusion of 2nd-order correction in IPT, which is 
considered in details in the next section.
\section{The improved perturbation theory(IPT) in NGAS}
In the context of NGAS, the IPT is the development of the standard Rayleigh-Schr\"{o}dinger perturbation series (RSPS) with the choice of the $unperturbed  Hamiltonian$ as the input-Hamiltonian, $H_0$ with the free-parameters in the latter being determined through PEQA and CVM (see,Eqs.(1-4)). The $perturbation, H^{'}$ is then defined by Eq.(6). The following properties of the IPT follow from the above choice of the Hamiltonian-splitting which are notable as being distinct from those of the conventional perturbation theory:\\(i) Using Eq.(7) and Eq.(16) it is seen that  the  perturbation-correction remains \emph{always} sub-dominant (by construction), compared to the unperturbed part in the following \emph{average-sense}:
\begin{equation}
|\langle~H_0\rangle|~~>>~|\langle~H^{'}\rangle|\equiv~~0.
\end{equation}
This property may be contrasted with the corresponding situation in the $conventional$ RSPS, e.g. for the case of the AHO where the `perturbation' ultimately prevails over the 'unperturbed' component of the Hamiltonian, no matter however small the quartic-coupling is.\\ 
(ii) Secondly, the 1st-order perturbation-correction identically vanishes due to Eq.(7),\\(iii) The IPT is \emph{not} restricted to small values of the coupling strength $\lambda$ since Eq.(37) holds for \emph{arbitrary} values of the latter.\\(iv) There is \emph{no} small-parameter naturally associated with the perturbation-however, the latter is small in the average sense as defined by Eq.(37). Therefore, to keep track of order-by-order corrections in IPT, one can adopt the standard trick of introducing an arbitrary, real but finite parameter, $\eta$ through the substitution: $H^{'}\rightarrow~{\eta}H^{'}$ and set this parameter to unity after the computation.\\(v)Property (iv) further implies  the $``universality''$ of application of IPT to \emph{arbitrary} interaction since the perturbation,$H^{'}$ does \emph{not} involve any parameter of the original Hamiltonian as the expansion-parameter for the RSPS. 

Because of the property (ii) the first non-trivial correction for the $n^{th}$ energy level starts at the 2nd-order and given by the standard expression:
\begin{equation}
\Delta{E_n^{(2)}}~=~\sum_{m\neq~n}\dfrac{\langle~n|H^{'}|~m\rangle\langle~m|H^{'}|~n\rangle}{E_n^{(0)}-E_m^{(0)}}
\end{equation}
In Eq.(38), we have used the notation:
\begin{equation}
\langle~n|H^{'}|~m\rangle~\equiv~\int_{-a(n,\lambda)}^{+a(n,\lambda)}dx~{\phi_n^{\ast}(x)~ H^{'}(n,\lambda)~\phi_m(x)}
\end{equation}
where $H^{'}$ is defined by Eq.(6) and we have displayed the $(n,\lambda)$-dependence of the relevant quantities. As an example, the perturbation-part of the Hamiltonian for the case of the quartic-AHO is given by 
\begin{equation}
{H^{'}|_{AHO}}~=~{\frac{1}{2}~gx^2}~+~{\lambda{x^4}}-{h(n,\lambda)},
\end{equation} 
where $h(n,\lambda)$ is given by Eq.(26). It may be noted that the RSPS can be derived \cite{ar9} \emph{without} invoking the properties of the  eigen-functions of the unperturbed Hamiltonian. Therefore, the non-orthogonality properties as expressed in Eq.(27) do not affect the standard formule of the RSPS.  

As has been discussed earlier and noted in the Tables-(I-III), the inclusion of the 2nd-order correction to the energy levels defined as: $ E_n^{(2)}\equiv~E_n^{LO}~+~\Delta{E_n^{(2)}}$ significantly improves the accuracy of the approximation in all the cases considered here, viz the AHO, DWO and the SHO. (Computation of still higher-order corrections to energy-levels falls beyond the scope of the present work but can be carried out by standard techniques in a straight-forward manner).

Finally, we summarize and discuss the main results of the present work in the next section.

\section{Summary and conclusions}
In summary, we have presented a very simple yet accurate approximation in the framework of the NGAS for the quartic anharmonic and the double-well oscillators using the elementary system of the infinite-square-well(ISW) as the approximating Hamiltonian. In the computation of the energy-spectrum, uniform accuracy is achieved to within a few percent of the exact numerical results \emph{even} in the leading-order, for \emph{arbitrary} values of the quartic-coupling, $\lambda$ and level index,$n_s$ for  \emph {all} the above systems. This situation may be contrasted with the results obtained using text-book methods of approximations, e.g. the naive perturbation method, the variational calculations, the WKB-method etc. Besides, the formalism \emph{naturally} reproduces the correct analytic-structure of the energy in the complex-$\lambda$ plane otherwise established through rigorous mathematical analysis. To systematically improve the leading-order results further, an improved perturbation theory is formulated in NGAS which is  \emph{not} restricted to small-coupling-expansion and shown to yield further significant improvement in accuracy with the inclusion of the 2nd-order correction only. The results and the method are directly relevant in a pedagogic-context owing to the extreme simplicity of the scheme (NGAS) as well as, the input ISW-approximation and further because it uses tools and techniques well within the grasp of a student-learner of a standard course of elementary quantum theory. The method can be extended in a straight forward manner to deal with systems in higher space-dimensions and to cases of higher anharmonicity.

\begin{acknowledgments}

This investigation was carried out at the National Institute of Science Education and Research (NISER), Bhubaneswar, India while one of the authors (BPM) served on the faculty as a visiting Professor. He gratefully acknowledges the facilities for research extended to him at the Institute.

\end{acknowledgments}

\begin{table}
    \caption{Results in the Leading order(LO) and with inclusion of $2^{nd}$ order perturbation-correction in IPT for the energy-levels of the quartic-AHO computed for sample values of $\lambda$, $n_s$ and for $g~=~1$ (see,Eq.(8)of text). Also shown, are corresponding results from some earlier computations (ref.7) denoted as $Exact$ under column-5.The entries in the last column and in the $3^{rd}$ column correspond to relative-$\%~ error$ with respect to the $`Exact'$ results.}
    \begin{ruledtabular}
        \begin{tabular}{c c c c c c c}
            $n_s$ & $\lambda$ & $E_n^{LO}$ &$Error(LO)$ & $E_n^{(2)}$ & $Exact$ & $Error$ \\ 
            &  &  & ${(\%)}$ & ref.(8) &   & ${(\%)}$\\
            \hline  
            0 & 0.1 & 0.6312 & 12.886 & 0.5748 & 0.5591  & 2.793\\ 
            & 1.0 & 0.9033 & 12.386 & 0.8290 & 0.8038 & 3.136 \\ 
            & 10.0 & 1.6902 & 12.311 & 1.5546 & 1.5049 & 3.296\\ 
            & 100.0 & 3.5168 & 12.310 & 3.2359 & 3.1314 & 3.341\\ 
            \hline 
            1 & 0.1 & 1.9636 & 10.972  & 1.8058 & 1.7695 & 2.055\\ 
            & 1.0 & 3.0366& 10.900 & 2.7999  & 2.7379 & 2.267\\ 
            & 10.0 & 5.9051 & 10.911 & 5.4479 & 5.3216 & 2.374\\ 
            & 100.0 & 12.4162 & 10.965 & 11.4561 & 11.1872 & 2.403\\ 
            \hline  
            2 & 0.1  &  3.3470 &  6.640 & 3.0943  & 3.1386 & 1.412 \\ 
            & 1.0 & 5.5818  & 7.771  & 5.1796  & 5.1793 & 0.006 \\ 
            & 10.0 & 11.9046  &  8.151& 10.3893 & 10.3471 & 0.408 \\ 
            & 100.0 & 23.7124 & 8.242& 22.0163 & 21.9069 & 0.499\\ 
            
            \hline  
            4 & 0.1& 6.3332 & 1.812 & 5.8162 & 6.2203 & 6.496 \\ 
            & 1.0 & 11.2748 &  2.634& 10.3198 & 10.9636 & 5.872 \\ 
            & 10.0 & 23.1124  &  2.838& 21.1642 & 22.4088 & 5.556\\ 
            & 100.0 & 49.2384 & 3.139 & 45.0952 & 47.7072 & 5.475 \\ 
            \hline  
            10 & 0.1& 16.8320 & 2.997  & 18.5203 & 17.3519 & 6.733\\
            & 1.0 & 32.1164 & 2.479 & 35.3936 & 32.9326 & 7.471\\ 
            & 10.0 & 67.1872 &  2.350 & 73.8641 & 68.8036 & 7.355 \\ 
            & 100.0 & 143.8104 & 2.323 & 157.9952 & 147.2267& 7.312\\

        \end{tabular}
    \end{ruledtabular} 
    \label{$Table-I$}
\end{table}
\begin{table}
    \caption{Results in the Leading order(LO) and with inclusion of $2^{nd}$ order perturbation-correction in IPT for the energy-levels of the Simple Harmonic Oscillator computed for sample values of $n_s$ and for $g~=~1$,(See,Eq.(28)of text). The entries in the 3rd- and the last column represent estimation of errors for the  results in the LO and with inclusion of 2nd-order correction, as compared to the `exact' results in the 5th-column.}    
    
    \begin{ruledtabular}
        \begin{tabular}{c c c c c c}
            $n_s$ & $E_n^{LO}$ & $Error-LO{(\%)}$ & $E_n^{(2)}$ & $Exact$ & $Error{(\%)}$ \\     
            \hline
            0 & 0.5678 & 13.56 & 0.5091 & 0.5 & 1.808 \\ 
            \hline
            1 & 1.6703 & 11.35 & 1.5239 & 1.5 & 1.578 \\ 
            \hline
            2 & 2.6272 & 5.05& 2.3888 & 2.5  & 4.462  \\ 
            \hline  
            5 & 5.3952 & 1.90 & 5.6456 & 5.5 & 2.609  \\ 
            \hline  
            10 & 9.9508 & 5.23 & 10.3945 & 10.5 & 1.155 \\
            \hline
            15 &  14.493 & 6.49 & 15.8155 & 15.5 & 1.260 \\
            
        \end{tabular}
    \end{ruledtabular} 
    \label{$Table-II$}
\end{table} 

\begin{table}
    \caption{Results in the Leading order(LO) and with inclusion of $2^{nd}$ order perturbation-correction in IPT for the energy-levels of the quartic-DWO computed for sample values of $\lambda$ and $n_s$ and for $g~=~1$ (see, Eq.(33) of text). Also shown, are corresponding results from ref.7 denoted as $Exact$ under column-6.The entries in the last column correspond to $({\%})error$ of results under column-5 with respect to those under column-6. The entries under column-3 represent relative accuracy of the LO-results as compared with the $Exact$ results shown under column-6.}    
    
    \begin{ruledtabular}
        \begin{tabular}{c c c c c c c}
            $n_s$ & $\lambda$ & $E_n^{~ref}$ & $Error(LO){(\%)}$ & $E_n^{(2)}$ & $Exact~(~ref.8)$ & $Error{(\%)}$ \\     
            \hline
            0 & 0.1 & 0.5049 & 7.220 &  0.4726 & 0.4709 & 0.350\\ 
            & 1.0 & 0.6422 &  11.242& 0.5967 & 0.5773 & 3.358\\ 
            & 10.0 & 1.5468 & 12.266 & 1.4246 &  1.3778 & 3.398\\ 
            & 100.0 & 3.4480 & 12.313 & 3.1732 & 3.0701 & 3.359\\ 
            \hline
            1 & 0.1 & 0.8252 & 7.476 &  0.7857  & 0.7678& 2.333\\ 
            & 1.0 & 2.3101 & 10.898 & 2.1366 &  2.0830 &  2.573\\ 
            & 10.0 & 5.5457 & 11.009 & 5.1180 & 4.9957 & 2.449 \\ 
            & 100.0 & 12.2471 & 10.993 & 11.3007 & 11.0337 & 2.419 \\ 
            \hline
            2 & 0.1  &  1.7393 & 6.392 & 1.6632 & 1.6348  & 1.740  \\ 
            & 1.0 & 4.6254 & 8.741 & 4.2983 & 4.2536  & 1.052   \\ 
            & 10.0 & 10.7242 &  8.378 & 9.9565 & 9.8947 & 0.625 \\ 
            & 100.0 & 23.4937 & 8.292& 21.8101 & 21.6947 & 0.532 \\ 
            \hline  
            4 & 0.1&  3.8678 & 4.994& 3.6471 & 3.6836 & 0.991   \\ 
            & 1.0 & 9.9139 & 3.659& 9.0975 & 9.5641 & 4.878  \\ 
            & 10.0 & 22.4582 &  3.322& 20.5659 & 21.7365 & 5.385 \\ 
            & 100.0 & 48.9325 & 3.248 & 44.7980 & 47.3929 & 5.475 \\ 
            \hline  
            10 & 0.1& 12.2697 & 1.292 & 12.840 & 12.4303 & 3.731 \\
            & 1.0 & 29.7738 & 2.139& 32.5902 & 30.4248 & 7.117  \\ 
            & 10.0 & 66.0787 & 2.327& 72.6644 & 67.6167 & 7.465 \\ 
            & 100.0 & 143.2937 & 2.361& 157.7026 & 146.6738 & 7.519 \\ 
            
        \end{tabular}
    \end{ruledtabular} 
    \label{$Table-III$}
\end{table}     
\end{document}